\documentclass[preprint,amsmath,superscriptaddress,pre]{revtex4}  
\usepackage[dvips]{graphicx}
\usepackage{epsfig}
\usepackage{color}
\usepackage{subfigure}
\usepackage{soul}

\begin{document} 

\title{
Scientific comment on 
``Tail risk of contagious diseases''
}
\author{\'Alvaro Corral}
\affiliation{%
Centre de Recerca Matem\`atica,
Edifici C, Campus Bellaterra,
E-08193 Barcelona, Spain
}\affiliation{Departament de Matem\`atiques,
Facultat de Ci\`encies,
Universitat Aut\`onoma de Barcelona,
E-08193 Barcelona, Spain
}\affiliation{Barcelona Graduate School of Mathematics, 
Edifici C, Campus Bellaterra,
E-08193 Barcelona, Spain
}\affiliation{Complexity Science Hub Vienna,
Josefst\"adter Stra$\beta$e 39,
1080 Vienna,
Austria
}
\begin{abstract} 
Cirillo and Taleb \cite{Cirillo_Taleb} study the size of major epidemics in human history in terms of the number of fatalities.
Using the figures from 72 epidemics, from the plague of Athens (429 BC)
to the COVID-19 (2019-2020), they claim that the resulting 
fatality distribution is ``extremely fat-tailed'', i.e., asymptotically a power law.
This has important consequences for risk, 
as the mean value of the fatality distribution becomes infinite.
%
%
%
Reanalyzing the same data 
we find that, although the data may be compatible with a power-law tail,
these results are not conclusive, 
and other distributions, not fat-tailed, could explain the data equally well.
Simulation of a log-normally distributed random variable
provides synthetic data whose statistics are undistinguishable from the statistics of the
empirical data.
\end{abstract} 
\maketitle

Cirillo and Taleb \cite{Cirillo_Taleb} identify ``fat-tailed'' distributions 
with regularly varying distributions \cite{Voitalov_krioukov}, 
defined by a complementary cumulative distribution function
(or survival function, probability of being above $x$) given by 
$S_{fat}(x) = \ell(x)/x^{\alpha}$, with $\alpha$ the exponent 
(of $S_{fat}(x)$, and $\xi=1/\alpha$ the tail index)
and $\ell(x)$ an unspecified slowly varying function
(for example, a function that tends to a constant when $x \rightarrow \infty$, 
but not only).
Roughly speaking, a ``fat-tailed'' distribution becomes a power law asymptotically.
Be aware that fat-tailed distributions are long-tailed distributions, which are heavy-tailed distributions in their turn, but not the opposite \cite{Voitalov_krioukov}.

%
%
%

As an alternative, we consider the truncated log-normal (ln) distribution,
which is not fat-tailed but is subexponential (and therefore long-tailed and heavy-tailed
\cite{Voitalov_krioukov}).
Its probability density is
$$
f_{ln}(x)=
{\sqrt{\frac 2\pi}}
\left[
 \mbox{erfc}\left(\frac{\ln u -\mu}{\sqrt{2} \sigma}\right)
\right]^{-1}
\frac 1{ \sigma x}
\exp\left(-\frac{(\ln x-\mu)^2}{2\sigma^2}\right),
$$
for $x\ge u$ (and zero otherwise), with $\mu$ and $\sigma$
the mean and standard deviation of the underlying (untruncated) normal distribution,
$u$ a lower cut-off,
and erfc the complementary error function. 
The log-normal distribution has been an important competitor of the power law
for the size distribution of structures and events in complex systems
\cite{Malevergne_Sornette_umpu,Corral_Gonzalez,Corral_Arcaute}.
%



\begin{figure}[!ht]
\begin{center}
(a)
\includegraphics[width=12.5cm]{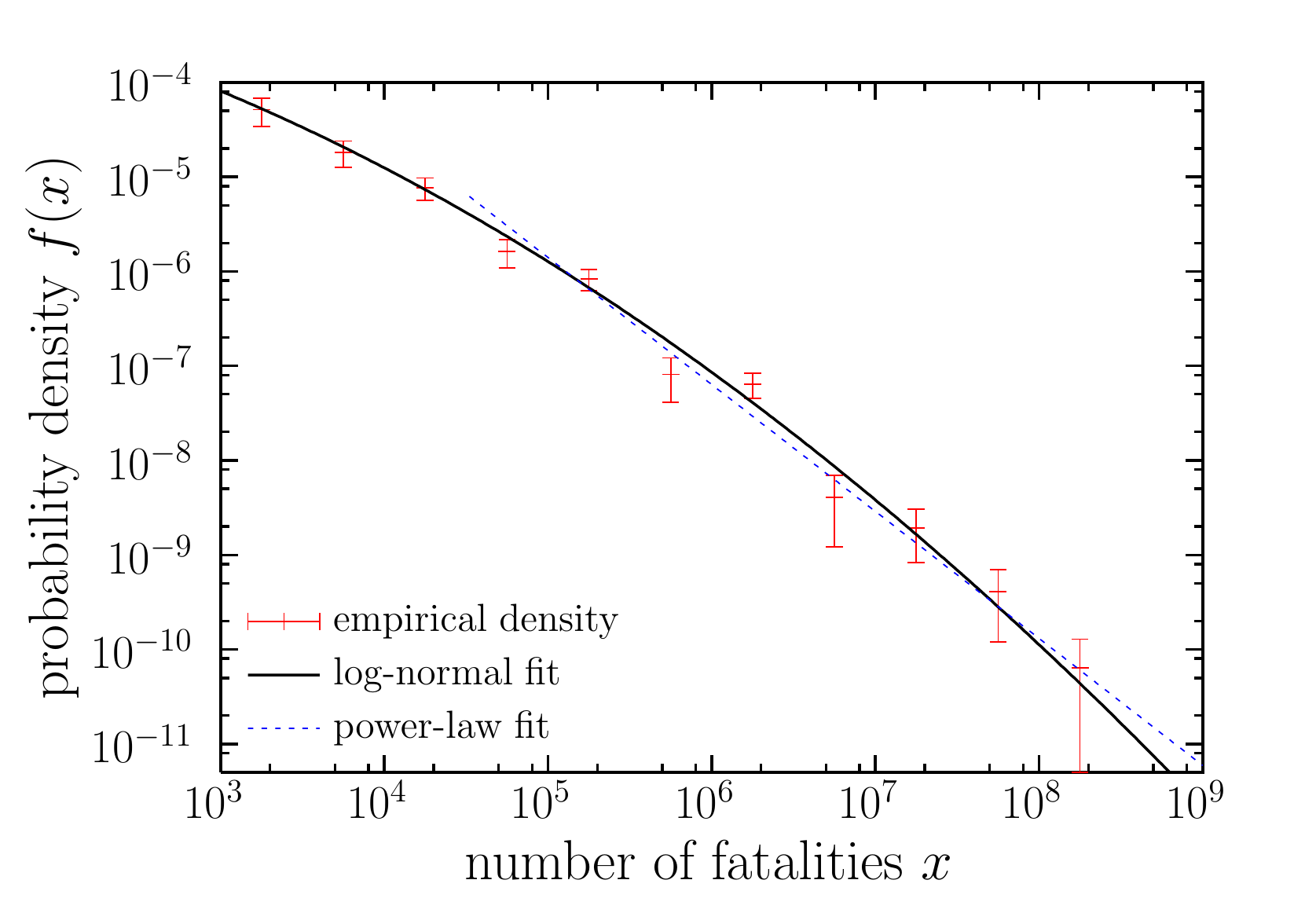}\\ 
(b)
\includegraphics[width=12.5cm]{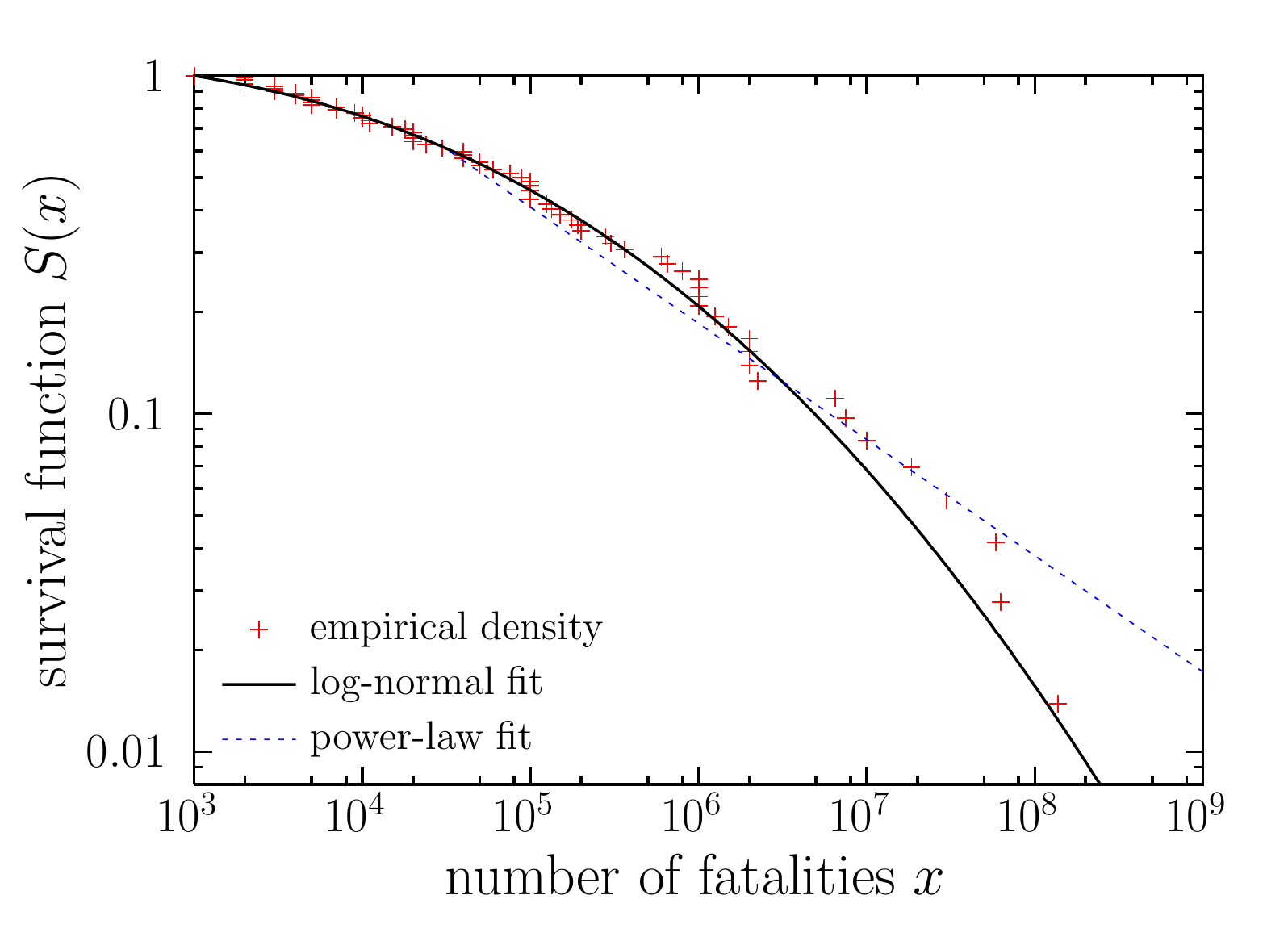} 
\end{center}
\caption{Empirical distribution
of the number of fatalities for each of the 72 historical epidemics
studied in Ref. \cite{Cirillo_Taleb}.
A truncated log-normal fit and a power-law tail (starting at $u\simeq 33,000$)
are shown as well.
(a) Probability density
 (empirical distribution
obtained using logarithmic binning \cite{Corral_Deluca}).
(b) Complementary cumulative distribution function
(survival function).}
\label{Fig2}
\end{figure}

We fit the truncated log-normal to the epidemic data of Ref. \cite{Cirillo_Taleb}
using the method of Ref. \cite{Corral_Gonzalez}, obtaining 
$u= 1000$ (fitting the whole data set),
$\mu=10.47$ 
and
$\sigma=3.58$
($p-$value $0.95$; scale parameter $e^\mu\simeq 35,200$).
The empirical estimation of the probability density of the data
together with the obtained fit
are shown in Fig. \ref{Fig2}.
A comparison 
with a power-law (pl) fit for the tail 
($f_{pl}(x) \propto 1/x^{1+\alpha}$, see Ref. \cite{Corral_Gonzalez})
shows that
both fits are very close to each other,
but also that the power law, with $u\simeq 33,000$ and $\alpha=0.344$
($p-$value $0.21$), 
gives a higher probability than the log-normal for the most extreme events (as expected).

Next we will compare the statistical behavior of the epidemic empirical data
of Ref. \cite{Cirillo_Taleb} with that of the simulation
of a truncated log-normal distribution (with the values of the parameters given above
and $N=72$ events).

%

\begin{figure}[!ht]
\begin{center}
(a)
\includegraphics[width=7.5cm]{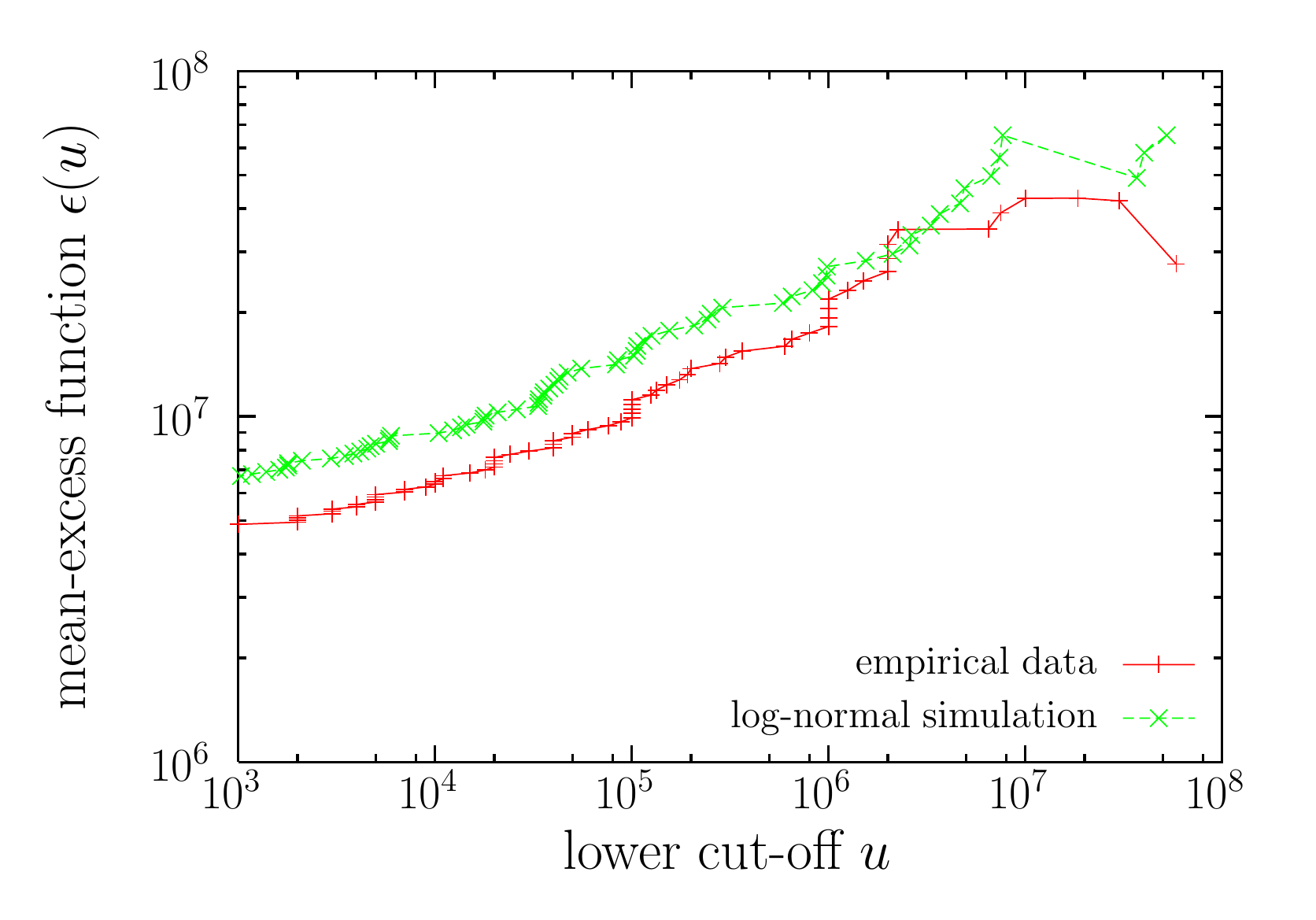}
(b)
\includegraphics[width=7.5cm]{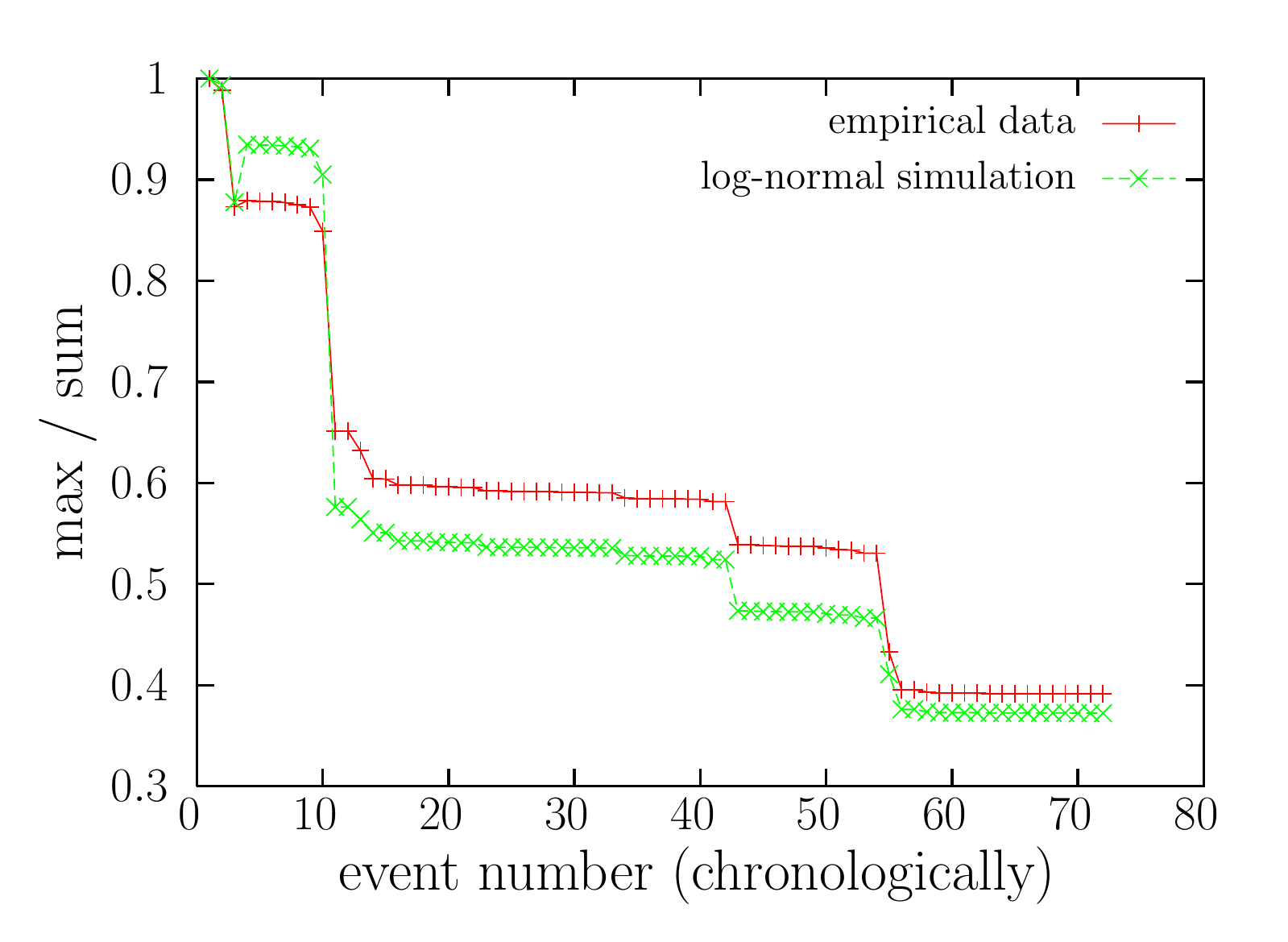}\\
(c)
\includegraphics[width=7.5cm]{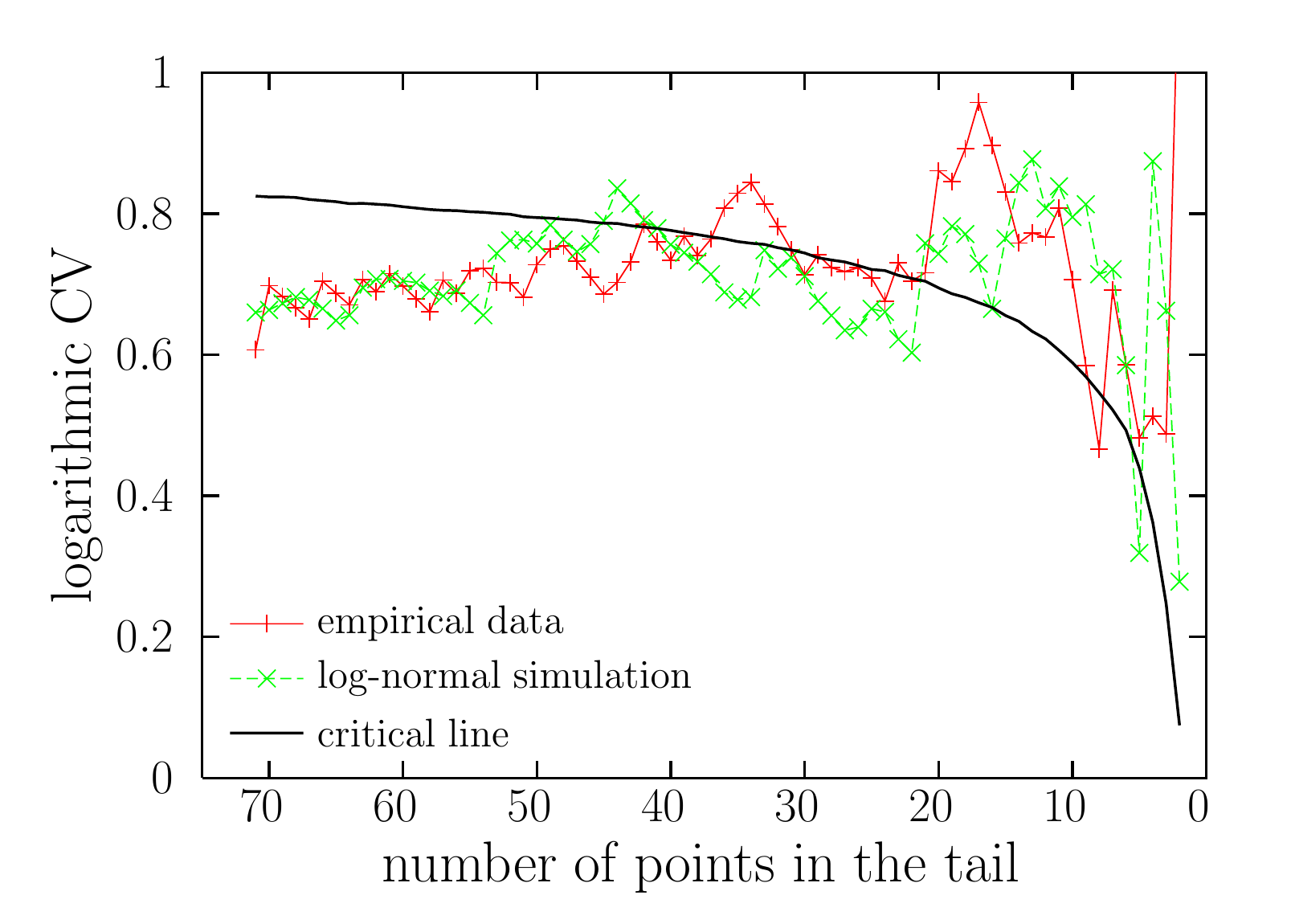}
(d)
\includegraphics[width=7.5cm]{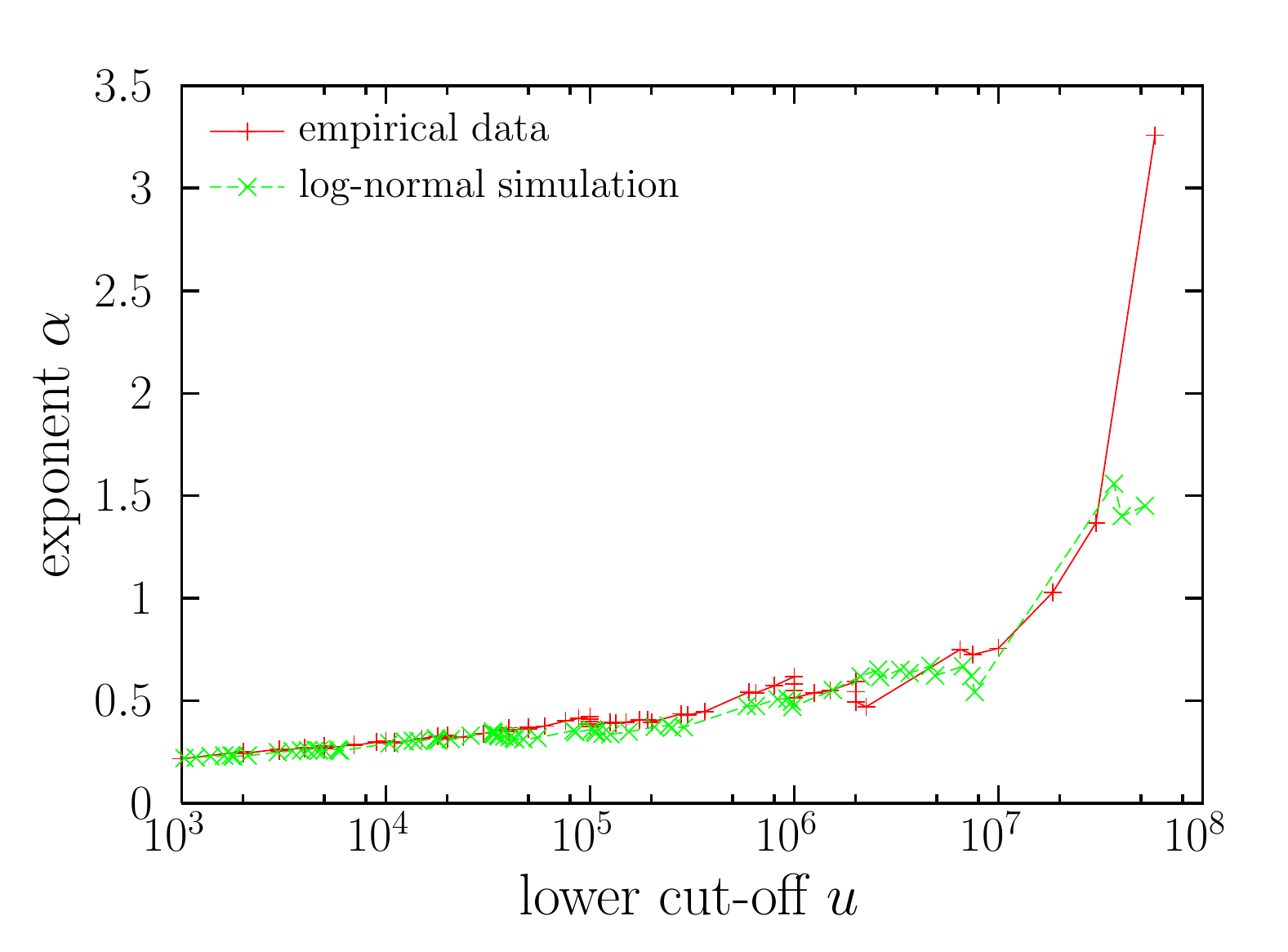}\\
\end{center}
\caption{
Statistical properties of the number of fatalities of historic epidemics
compared to those of a log-normal synthetic sample fitting the empirical data.
(a) Mean-excess function $\epsilon(u)$ versus minimum size (lower cut-off) $u$.
(b) Maximum-to-sum ratio as a function of number of data, 
in chronological order.
(c) Logarithmic coefficient of variation as a function of the number of points in the tail
(those with $x > u$; note that the horizontal axis is reversed).
(d) Power-law exponent $\alpha$ as a function of $u$.
}
\label{Fig5}
\end{figure}

Cirillo and Taleb \cite{Cirillo_Taleb} propose two main ways to check fat-tailness.
One of this uses
the mean-excess function $\epsilon(u)=\langle x-u | x\ge u \rangle$, 
where the brackets denote expected value.
This is the same as the expected residual size
\cite{Kalbfleisch2} (see also Ref. \cite{Schroeder}) 
used in reliability theory
and characterizing the distribution in a way totally equivalent
to $f(x)$ or $S(x)$, provided that the first moment of the distribution is finite
($\epsilon(0)=\langle x \rangle$, 
but note that for $\alpha < 1$ this is not the case).
Figure \ref{Fig5}(a) shows how
the mean-excess function of the log-normally simulated data 
shows the same pattern as the
result for the empirical data.

The other approach in Ref. \cite{Cirillo_Taleb} uses the (partial) maximum-to-sum ratio
(the maximum of the $x-$values divided by the sum of the values).
As $N\rightarrow \infty$, this ratio should tend to zero when the mean of the distribution
is finite (as it happens with the log-normal but not with the power law when $\alpha < 1$). We again compare the empirical data with the simulated data, 
sorting the simulated data in order that the ranks of the sizes 
(number of fatalities)
follow the same temporal pattern as the empirical data 
(i.e., the largest simulated event is always put on the 11th position, where the Black Death,
the largest event on record, takes place in the original data, and so on).
The results, displayed at Fig. \ref{Fig5}(b), 
show again that the behavior of synthetic log-normal data
is very close to that of the empirical data.
Thus, although the theory teaches us that the maximum-to-sum ratio tends to zero 
when $N\rightarrow \infty$ if the distribution has a finite mean, 
this convergence can be rather slow, 
as it happens with the log-normal distribution for the parameter values
that describe the epidemic data.

 


Now we provide complementary evidence that the log-normal distribution 
is a good fit of the epidemic data of Ref. \cite{Cirillo_Taleb}.
In fact,
the power law can be considered a
particular case of the truncated log-normal
(in the same way that the exponential is a particular case of the truncated normal distribution when $\sigma^2 \rightarrow \infty$ and $\mu \rightarrow -\infty$
\cite{Castillo}). 
In this sense, the log-normal will always provide a better fit.
However, on the other hand, 
it may happen that this improvement in the fit is not significant,
and then the power-law fit suffices for describing the data. 
This is something that can be evaluated by a likelihood-ratio (LR) test \cite{pawitan2001}.

Taking advantage of the fact that the LR between both distributions
is a decreasing function of the logarithmic coefficient of variation (CV) \cite{Castillo},
this provides a very simple way to perform the LR test
(without the need of performing maximum-likelihood fitting):
critical values of the LR translate into critical values of the logarithmic CV.
When this quantity is close enough to one, the power-law hypothesis cannot be rejected, 
and when it departs significantly from one (from below),
the power law is rejected in favor of the log-normal.  

The test is performed for different values of the lower cut-off $u$, 
and the results, for the complete data set, are displayed on Fig. \ref{Fig5}(c).
This shows that only for the 21 largest epidemics 
the power-law tail is not rejected in favor of the log-normal.
The corresponding cut-off $u$ turns out to be at about 350,000 fatalities.
In other words, the 21 epidemics with more than 350,000 casualties are well described
by a power law (the improvement brought by the log-normal is not significant), 
but, including events below 350,000, the log-normal fit is significantly better
(for the full range).
Applying the same procedure to the log-normally simulated data
reproduces again the pattern obtained for the empirical data, 
as also shown in the figure.

Assuming that a power law can describe the largest epidemics (in terms of fatalities),
which would be the value of the corresponding power-law exponent $\alpha$?
Above, for $u\simeq 33,000$, we report $\alpha=0.344$,
but for $u\simeq 350,000$ the value is larger.
In fact, the value of $\alpha$ is not stable at all, 
growing when the lower cut-off $u$ increases 
(this is
already apparent in the results of Cirillo and Taleb \cite{Cirillo_Taleb}), 
which prevents that one can establish a well-defined exponent
\cite{Baro_Vives}.
%
Indeed, Fig. \ref{Fig5}(d) shows the resulting exponents $\alpha$ as a function of $u$,
comparing the original (empirical) data with the log-normally simulated ones.
It is clear that the simulated data provides a pattern very similar to the empirical one, with an increase of the value of the exponent $\alpha$
when $u$ increases. 
Indeed, this increasing behavior of the fitted exponent is what one expects from 
a log-normal distribution.

We have shown how the probability distribution of the number of fatalities of historical epidemics can be well explained by a log-normal distribution, 
which is a distribution that is empirically similar to the power law
but quite different from a theoretical point of view (in particular, 
the mean and all moments of a log-normal distribution are well defined).
Our work shows the importance of considering alternative probability models
when fitting heavy-tailed distributed data (which is different from ``fat-tailed' data'' \cite{Voitalov_krioukov}), 
as well of the key role of computer simulations to contrast the validity 
of theoretical results when the number of data is not infinite.




I acknowledge discussions with Isabel Serra and
support from projects
FIS2015-71851-P and
PGC-FIS2018-099629-B-I00
from Spanish MINECO and MICINN.


\end{document}